\documentstyle[sprocl,epsfig,amssymb]{article}

\newcommand{\cF}{{\cal F}}

\newcommand{\EQ}{\begin{equation}}
\newcommand{\EEQ}{\end{equation}}
\newcommand{\EA}{\begin{eqnarray}}
\newcommand{\EEA}{\end{eqnarray}}

\newcommand{\ol}{\overline}

\newcommand{\ds}{\delta \sigma}
\begin{document}

\title{OPTIMIZING SMITH-WATERMAN ALIGNMENTS}

\author{ROLF OLSEN, TERENCE HWA}
\address{
        Department of Physics,
        University of California at San Diego\\
        La Jolla, CA 92093-0319\\
        email: rolf@cezanne.ucsd.edu, hwa@ucsd.edu
	}

\author{MICHAEL L\"ASSIG}
\address{
        Max-Planck-Institut f\"ur Kolloid- und
        Grenzfl\"achenforschung \\
        Kantstr.~55, 14513 Teltow, Germany \\
        email: lassig@mpikg-teltow.mpg.de \\
        \hspace*{0cm} }

\maketitle\abstracts{
Mutual correlation between segments of DNA or protein sequences can be detected
by Smith-Waterman local alignments. We present a  statistical
analysis of alignment of such sequences, based on a recent scaling theory. 
A new {\em fidelity} measure is introduced and shown to capture the  
significance of the local alignment, i.e., the extent to which  the correlated 
subsequences are correctly identified. It is demonstrated how the fidelity may 
be optimized in the space of penalty parameters using only the alignment score 
data of a {\em single} sequence pair.}

\section{Introduction}

Sequence alignment has become an indispensable tool in molecular
biology~\cite{Doolittle96}.
A number of different algorithms are available to date, and their variety
and complexity continues to grow~\cite{Waterman94}. For a given application,
however,
a suitable type of algorithm and optimal scoring parameters are still
chosen mostly on an empirical basis~\cite{Vingron94,Waterman94c,Altschul96}.
The practical problems in the
application of alignment algorithms reflect a number of
poorly understood conceptual issues:
Given sequences with mutual
correlations, how can the {\em fidelity} of an alignment --- i.e.,
the correlations correctly captured --- be quantified? How can the scoring
parameters be chosen to produce high-fidelity alignments? Are the results
statistically and biologically significant?

In a series of recent
publications~\cite{Hwa96,Hwa98,Drasdo98b,Drasdo98}, we have developed a
{\em statistical scaling theory} of gapped alignment aimed at addressing
these issues. This theory describes the dependence of alignment data on the
inter-sequence correlations and on the scoring parameters used. The entire
parameter dependence of alignments is contained in a number of
{\em characteristic scales}. For Smith-Waterman alignments~\cite{Smith81},
the most important scales are the typical
length $t_0$ of mutually uncorrelated subsequences locally aligned, and the
minimum length $t_c$ of mutually correlated subsequences detectable by
alignment. Expressed in terms of these characteristic scales,
the alignment statistics acquires
{\em universal} properties independent of the scoring parameters.
Hence, optimizing alignments reduces to optimizing the values of the
characteristic scales.

In this paper, we study the statistics of Smith-Waterman alignments for
piecewise correlated sequences. We define a suitable fidelity function
weighing appropriately aligned pairs of correlated elements against false
positives. The parameter dependence of the fidelity is found to be
captured by the scaling theory of alignment. High-fidelity alignments are
obtained if the characteristic
scales $t_0$ and $t_c$ are of the same order of magnitude and are jointly
optimized. For a given sequence pair, we show how this optimization
can be obtained directly from the score data, leading to the central result of this
paper: a simple procedure for optimizing the fidelity of  Smith-Waterman alignments.

\section{The Smith-Waterman Algorithm}

We study local alignments of pairs of Markov sequences
$Q = \{Q_i\}$ and
$Q'= \{Q'_j \}$ with an approximately equal number of elements $\sim N/2$.
Each element $Q_i$ or $Q'_j$
is chosen with equal probability from a set of $c$ different letters,
independently of the other elements of the same sequence. There may, however,
be inter-sequence correlations in pairs $(Q_i, Q_j')$.
We here take $c = 4$, as is appropriate for nucleotide sequences, although
the results can be easily generalized to arbitrary values of $c$. An alignment
is defined as an ordered set of pairings $(Q_i, Q'_j)$ and of gaps
$(Q_i,-)$ and $(-,Q'_j)$ involving the elements of two contiguous
subsequences $\{Q_{i_1}, \dots, Q_{i_2} \}$ and
$\{ Q'_{j_1}, \dots, Q'_{j_2} \}$; see Fig.~1(a).
We define the length of an alignment as the total number of aligned
elements of both sequences, $L \equiv  i_2 - i_1 + j_2 - j_1$.

A given alignment is conveniently  represented~\cite{Needleman70}
as a {\em directed path} on a two-dimensional grid as shown in Fig.~1(b).
Using the rotated coordinates $r\equiv i-j$ and $t\equiv i+j$, this path is
described by a single-valued function $r(t)$ measuring the ``displacement''
of the path from the diagonal of the alignment grid. The length $L$ of
the alignment equals the projected length of its path onto the diagonal.

\begin{figure}[t]
\centerline{\epsfig{file=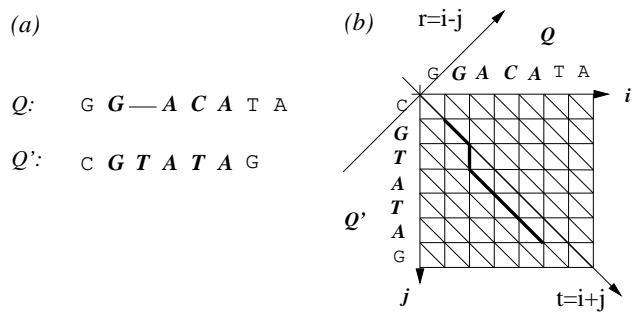, height=1.6in, angle=0}}
\caption{\small
(a) One possible local alignment of two nucleotide sequences $Q$ and $Q'$.
The aligned subsequences are shown
in boldface, with $4$ pairings (three matches, one mismatch)  and one gap.
The alignment contains a total of $L = 9$  elements.
(b) Unique representation of the alignment in (a)
as  {\em directed} path $r(t)$ (the thick line) on a two-dimensional
alignment grid. Each vertical (horizontal) bond of the path corresponds
to a gap in sequence $Q$ ($Q'$).  $L$ is
the projected length onto the $t$ axis.
}
\end{figure}

Each alignment is assigned a
score $S$, maximization of which defines the optimal alignment for a
given scoring function.
The simplest class of {\em linear} scoring functions is of the form
$S = \sigma_+ N_+ + \sigma_- N_- + \sigma_g N_g$, where
$N_{+}$ is the total number of matches
$(Q_i = Q'_j)$, $N_{-}$ the  number of mismatches
$(Q_i \neq Q'_j)$, $N_g$ the number of gaps, and
$\sigma_+$, $\sigma_-$, $\sigma_g$ are the associated scoring parameters.
Since an overall multiplication of the score does not change the
alignment result, we can use the normalized scoring function
\EQ
S = \sigma \, L +
    \sqrt{c - 1}\, N_+ - \frac{1}{\sqrt{c - 1}} \, N_- - \gamma N_g
\label{S}
\EEQ
with $L = 2N_+ + 2N_- + N_g$ denoting again the alignment length
defined above. This form of the scoring function contains the
two natural scoring parameters: the score gain $\sigma$ per aligned element,
and the gap cost $\gamma$. The parameter $\sigma$ controls the length $L$ of
the optimal alignment; while changing $\gamma$ affects its number of gaps,
 i.e., the mean square displacement of the optimal alignment path
from the diagonal of the alignment grid. (Borrowing
notions from physics and chemistry, we can think of the alignment path $r(t)$
as a {\em polymer} stretched along the $t$ axis, with ``chemical potential''
$\sigma$ and ``line tension'' $\gamma$.)

We use the Smith-Waterman recursion relation~\cite{Smith81}
\begin{equation}
S(r,t) = \max \left \{ \begin{array}{l}
                       S(r-1, t-1) + \sigma - \gamma \\
                       S(r+1, t-1) + \sigma - \gamma \\
                       S(r,   t-2) + s(r,t) + 2\sigma \\
                       0
                       \end{array} \right \}
\label{tm}
\end{equation}
with
\begin{equation}
s(r,t) = \left \{ \begin{array}{lll}
                  \sqrt{c - 1} \;\;\;& \mbox{ if } &
                  Q_{(r+t)/2} = Q'_{(t-r)/2} \\
                  - \frac{1}{\sqrt{c-1}} \;\;\;& \mbox{ if } &
                  Q_{(r+t)/2} \neq Q'_{(t-r)/2}
                  \end{array}  \right. \;
\end{equation}
and suitable boundary conditions~\cite{Drasdo98}.  $S(r,t)$ is the score
maximum  for the set of all alignment paths ending at the point $(r,t)$. The
optimal alignment ends at the point $(r_2, t_2)$ defined by the global score
maximum, $S(r_2,t_2) = \max_{r,t} S(r,t)$. The entire path is then traced back
from the endpoint to the initial point $(r_1,t_1)$ given by $S(r_1,t_1) = 0$.
The length of the  optimal path is $L=t_2-t_1$. For large values of $\sigma$,
the  optimal alignment of long sequences becomes a so-called {\em global}
alignment  involving the entire sequences $Q$ and $Q'$ up to small unpaired
regions at both ends; i.e., $L \simeq N$. In this limit, the  Smith-Waterman
algorithm becomes equivalent to the simpler  Needleman-Wunsch
algorithm~\cite{Needleman70}.

\section{Scaling of Smith-Waterman alignments}

The statistical theory of alignment describes averages (denoted
by overbars) over an {\em ensemble} of sequence pairs with well-defined
mutual correlations. However, we emphasize that the properties of {\em single}
pairs of ``typical'' sequences are well approximated by these
averages~\cite{Hwa98}.

The simplest form of scaling is realized
in the limit of global alignment ($\sigma \to \infty$)
for pairs of Markov sequences without mutual correlations.
Important statistical averages then scale as powers of the sequence length;
for example, the variance of the optimal score
$\overline{(\Delta S)^2} \propto N^{2/3}$. The exponents of
these power laws are {\em universal}, i.e., independent of the scoring
parameters. A detailed discussion was given
by Drasdo {\it et al}~~\cite{Drasdo98}.

For generic values of $\sigma$, the alignment statistics becomes more
complicated even for mutually uncorrelated sequences. Most importantly,
there is a phase transition~\cite{Arratia94} along a critical line
$\sigma = \sigma_c (\gamma)$. For $\sigma > \sigma_c$, the optimal
alignment of long sequences remains global; i.e., it has asymptotic length
$L \simeq N$ and score $S \propto N$ for $N \gg 1$. This is called
the {\em linear phase}. For $\sigma < \sigma_c$,
however, the optimal alignment ending at a given point $(r,t)$ remains
finite. The limit values of its average length and score,
$t_0 \equiv \lim_{t \to \infty} \ol L(t)$ and
$S_0 \equiv \lim_{t \to \infty} \ol S(t) $, are {\em characteristic scales}
asymptotically independent of the sequence length $N$. (The argument $r$
has been suppressed since these averages are independent of it.) The global
optimal alignment path is then of length $L \sim t_0 \log N$, 
which gives the name
{\em logarithmic phase} to the regime $\sigma < \sigma_c$.

Close to the phase transition, the characteristic scales themselves
diverge as powers of the distance
$\ds \equiv \sigma - \sigma_c (\gamma)$ to the critical line~\cite{Drasdo98b},
\begin{equation}
t_0(\sigma,\gamma) \sim B^{3/2}(\gamma) |\ds|^{-3/2}
\; , \;\;\;
S_0(\sigma,\gamma) \sim B^{3/2}(\gamma) |\ds|^{-1/2} \;.
\label{tsSsat}
\end{equation}
(Here $\sim$ denotes proportionality with a $(\sigma,\gamma)$-independent
proportionality constant.)
The coefficient function $B(\gamma)$ and the critical line $\sigma_c (\gamma)$
are known numerically~\cite{Drasdo98,Drasdo98b}.
In this region, the average length and score take the scaling
form~\cite{Hwa98}
\begin{equation}
\frac{\ol{L}(t)}{t_0} = {\cal L}_\pm \! \left ( \frac{t}{t_0} \right )
\;, \;\;\;
\frac{ \ol{S} (t)}{S_0}
                      = {\cal S}_\pm \! \left ( \frac{t}{t_0} \right ) \;.
\label{calS}
\end{equation}
The subscript of the scaling functions ${\cal L}$ and ${\cal S}$ 
refers to the sign of $\ds$; the
two branches correspond to the linear and the logarithmic phase, respectively.
The entire dependence on the scoring parameters is contained
in the characteristic scales (\ref{tsSsat}), while the scaling functions
${\cal S}_\pm$ and ${\cal L}_\pm$ are again universal.
The meaning of the scaling form (\ref{calS}) is quite simple: It relates
alignment data for different values of the scoring parameters.
This leads to the data collapse of Fig.~2.
\begin{figure}[tbh]
\begin{center}
\epsfig{file=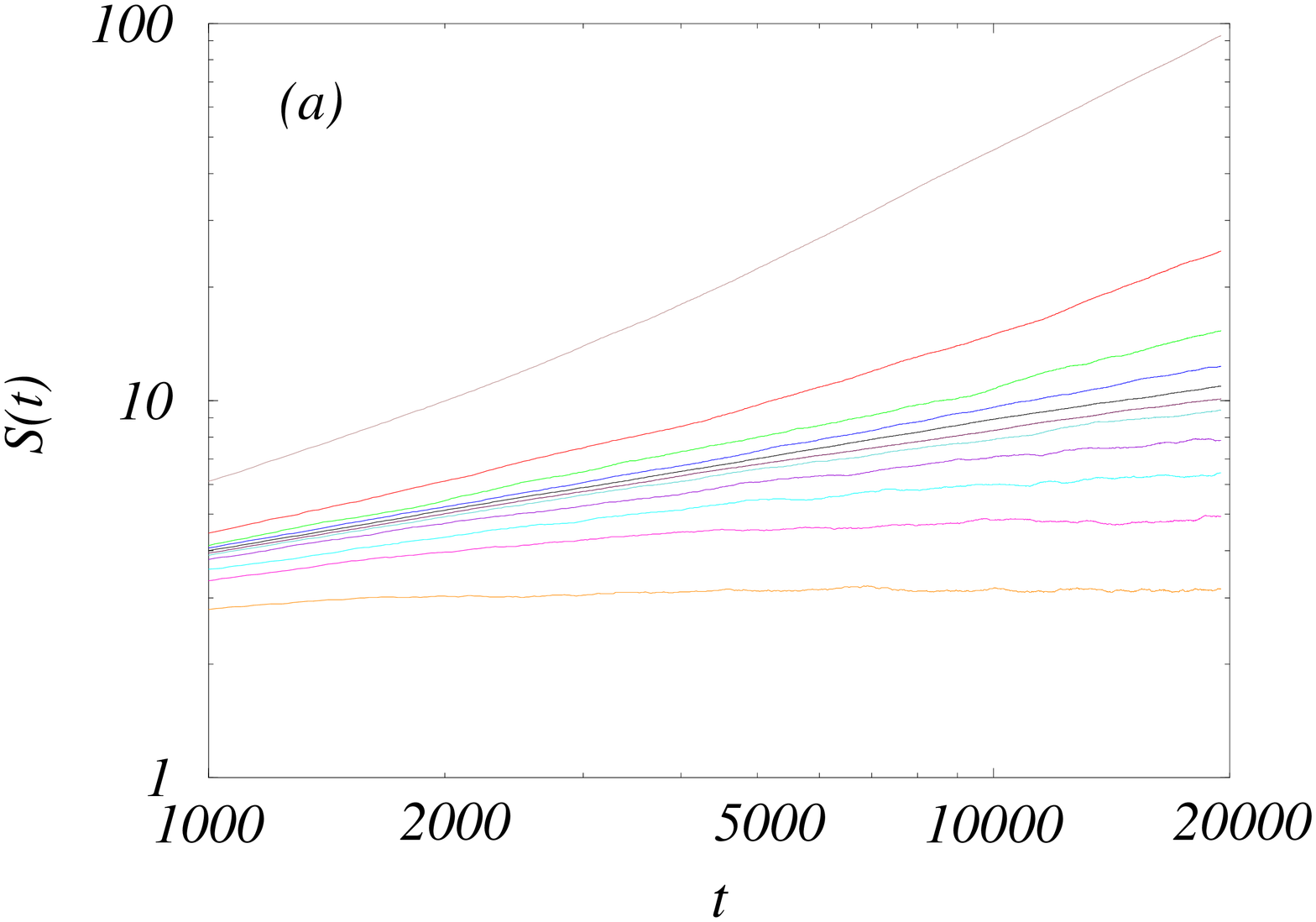, height=2.55in, angle=270}
\epsfig{file=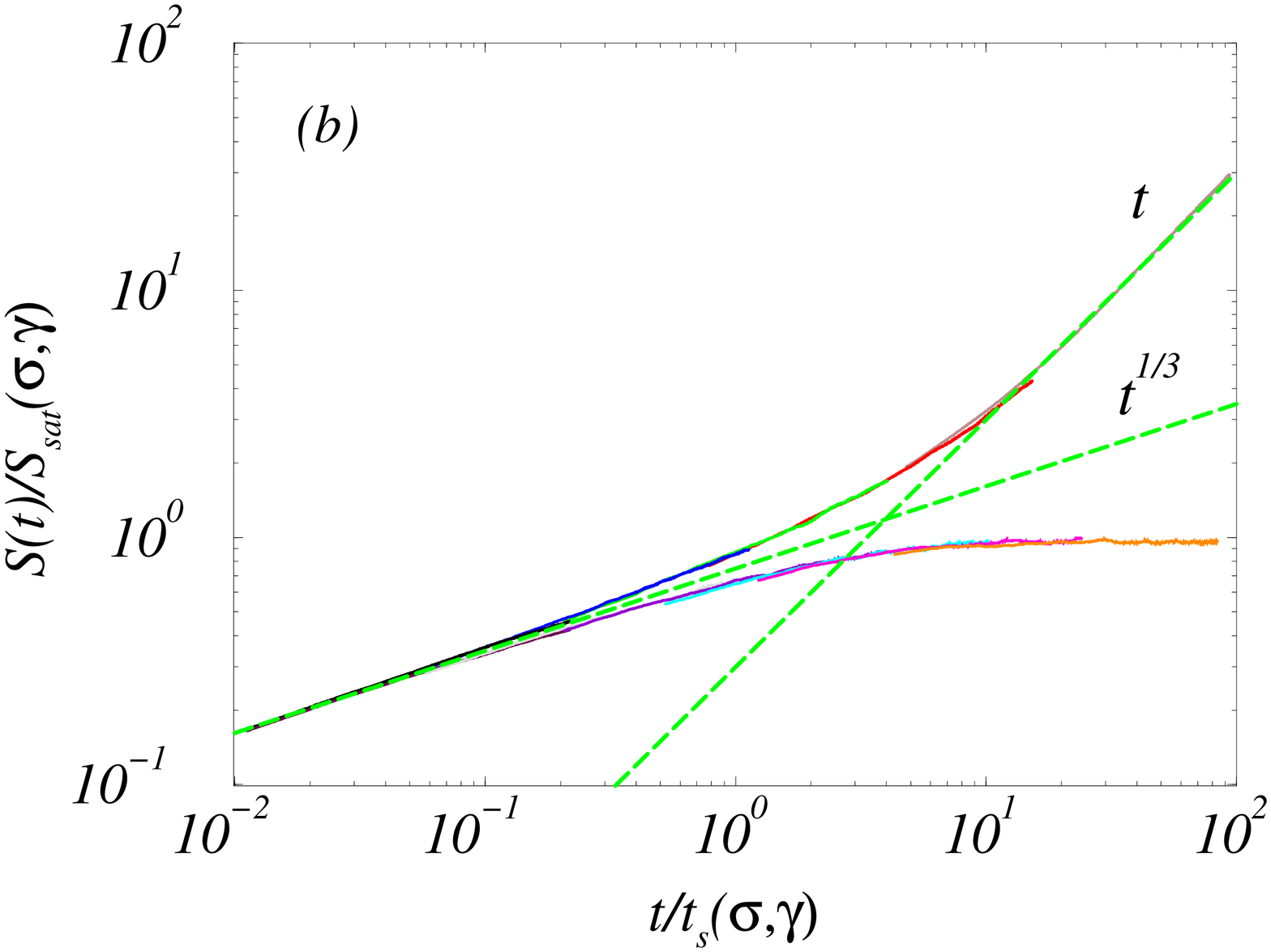, height=2.55in, angle=270}
\end{center}
\vspace{\baselineskip}
{\small Fig. 2: Local alignment of sequences without
mutual correlations.
(a) Average score $\ol S (t)$ (over an ensemble of 1000 random sequence
pairs of $10000$ elements each) of the optimal alignment for various
scoring parameters. The curves correspond to $\gamma = 3.0$ and $
\delta \sigma / \sigma_c(\gamma) = 0.05$ to $-0.05$ (top to bottom).
(b) The scaled curves $\ol S(t) / S_0$ as  functions of
$t/t_0$ collapse to the universal two-branched function ${\cal S}_\pm$ of
Eq.~(\ref{calS}). The asymptotics of this function is given by
power laws (dashed lines) predicted by the scaling theory~\cite{Hwa98}.}
\end{figure}

We now turn to alignments of Markov
sequences $Q$ and $Q'$ with {\em mutually correlated} subsequences
$\hat Q$ and $\hat Q'$ (referred to below as {\em target}) of approximately
equal length $\hat{N}/2$.
The ``daughter'' sequence $\hat Q'$ is obtained from the ``ancestor''
sequence $\hat Q$ by a simple Markov evolution process~\cite{Drasdo98} with
substitution probability $p$ and insertion/deletion probability $q$.
The average fraction $U = (1 - p)(1 - q)$ of ancestor elements conserved in
the daughter sequence quantifies the degree of correlations between
$\hat Q$ and $\hat Q'$. The remainder of $Q$ and $Q'$ has no correlations.

A meaningful alignment of the sequences $Q$ and $Q'$ should (i)~match a
fair fraction $f$ of the pairs of conserved elements
$(Q_i, Q'_j) \in \hat Q \times \hat Q'$ and (ii)~remain confined to
the target region to avoid false matches. We quantify these properties by the
 {\em fidelity function}
\begin{equation}
\cF =  \frac{2 \hat N}{L + \hat N} \, f \;, \label{fidelity}
\end{equation}
which takes values between 0 and 1. The prefactor is designed to penalize
local alignments that are too long ($L > \hat N$). Its precise form
influences the parameter dependence of the fidelity only weakly.
For global alignments, $\cF$ reduces to the fidelity function used
previously~\cite{Drasdo98}, $\cF = f$. Maximizing $\cF$ for a given pair of
sequences should produce an alignment of {\em bona fide} biological
significance.

Alignments of correlated sequences have a second set of characteristic scales
related to their statistical significance~\cite{Drasdo98}.
The {\em threshold} or
{\em correlation} length $t_c(\gamma)$ is the minimal length
of a target to be detectable statistically  by alignment~\footnote{
More precisely, we consider global alignments ($\sigma \to \infty$)
of sequences of length $N$ with mutual correlations over their entire length
(i.e., $\hat Q = Q$ and $\hat Q' = Q'$). 
For $\hat N < t_c$, however,
random agglomeration of matches outweigh the pairs of correlated elements,
rendering the correlation undetectable. See Ref.~\cite{Drasdo98} for details.}.
($t_c$ also depends on the
evolution parameters, in the present case $U$ and $q$, but is independent
of $\sigma$.)
In the sequel, we study targets
of length $\hat N$ well above $t_c$ and well below the overall length $N$.
The relevant ensemble averages can then
again be written in scaling form. For the fidelity and the length of
the optimal alignment, we expect the approximate expressions
\begin{equation}
\frac{\ol \cF}{\cF^* (\gamma)} =
                          \varphi     \left ( \frac{t_c}{t_0} \right )
\;, \;\;\;
\frac{\ol L}{\hat N} =    {\cal L} \left ( \frac{t_c}{t_0} \right )
\;,
\label{scalcor}
\end{equation}
where $\cF^* (\gamma) \equiv \max_\sigma \ol \cF (\sigma, \gamma)$ denotes the
relative fidelity maximum at a given value of $\gamma$.
The important point of this scaling form is again quite simple: It relates
alignment data at different values of the alignment parameters and of the
evolution parameters.  The scaling functions
$\varphi$ and $\cal L$ are universal as before, only their arguments
$t_c / t_0$ depend on the parameters.
This is crucial for finding optimal alignment parameters as we
show in the next Section.

The form of Eq.~(\ref{scalcor}) has been verified numerically:
Figs.~3(a) and~4(a) show the average fidelity and length of optimal alignments,
respectively, for different values of $\gamma$ and $\sigma$.
 The data for different parameter values are indeed related
as is evident from the collapse of the scaled curves 
$\ol \cF/\cF^*(\gamma)$ and $\ol L/\hat N$; see Figs.~3(b) and~4(b).
The scaled abscissa $\ds/ |\ds^*(\gamma)|$ can be expressed in terms of the
ratio of characteristic scales in (\ref{scalcor}),
$\ds/ |\ds^*(\gamma)| = (t_c / t_0)^{2/3}$, as follows from (\ref{tsSsat}) and
the relation $t_c(\gamma) \sim (\ds^*(\gamma)/ B(\gamma))^{-3/2}$ which is
anticipated from a previous analysis~\cite{Drasdo98b}.
Here, $\ds^* \equiv \sigma^*(\gamma) - \sigma_c(\gamma)$,
and $\sigma^*$ is the location of the relative fidelity maximum,
defined from $\cF^*(\gamma)=\overline{\cF}(\sigma^*,\gamma)$.
The data collapse shown in Figs.~3 and 4 therefore supports 
the proposed scaling form~(\ref{scalcor}).

\begin{figure}
\begin{center}
\epsfig{file=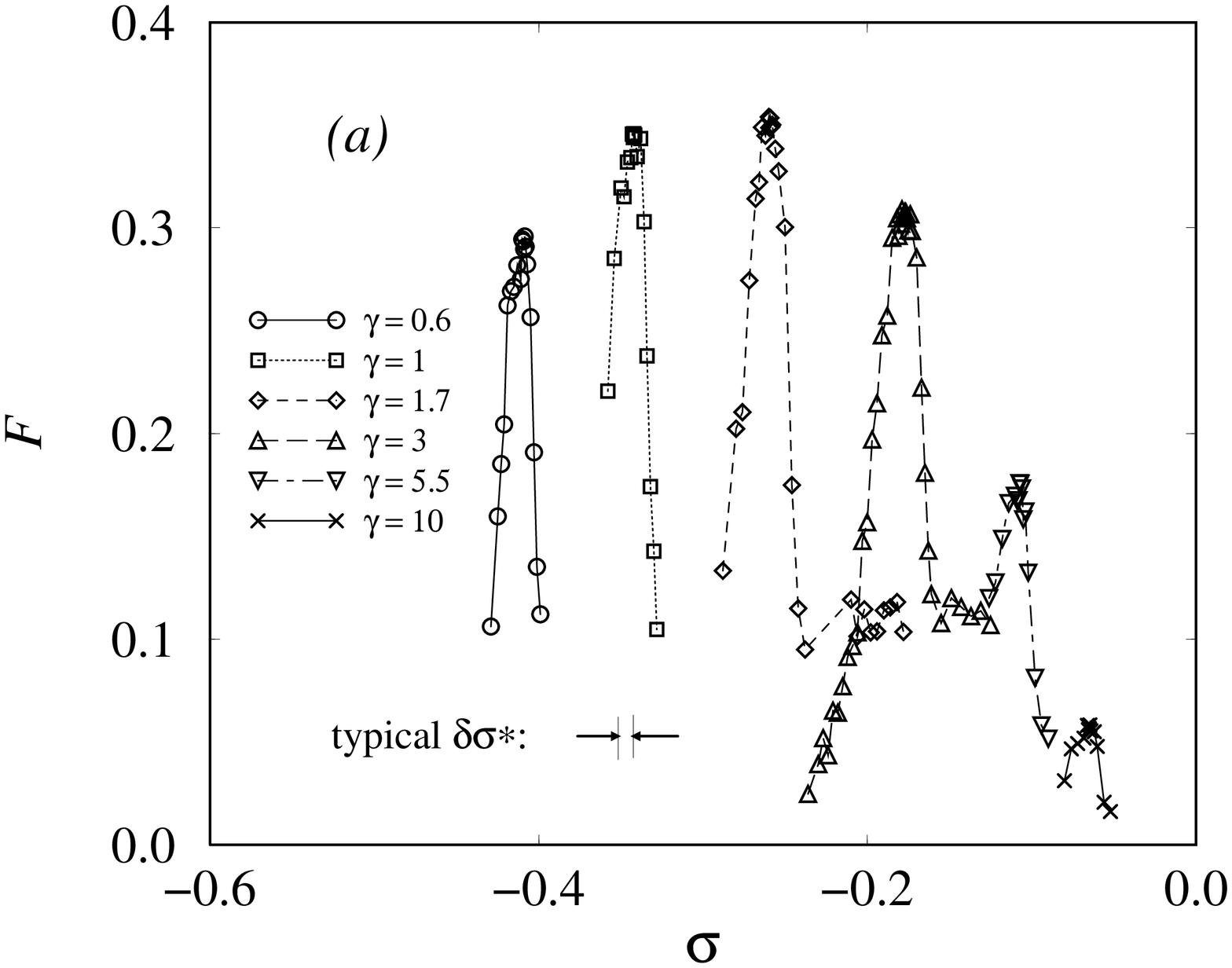, height=3.0in, angle=270}
\epsfig{file=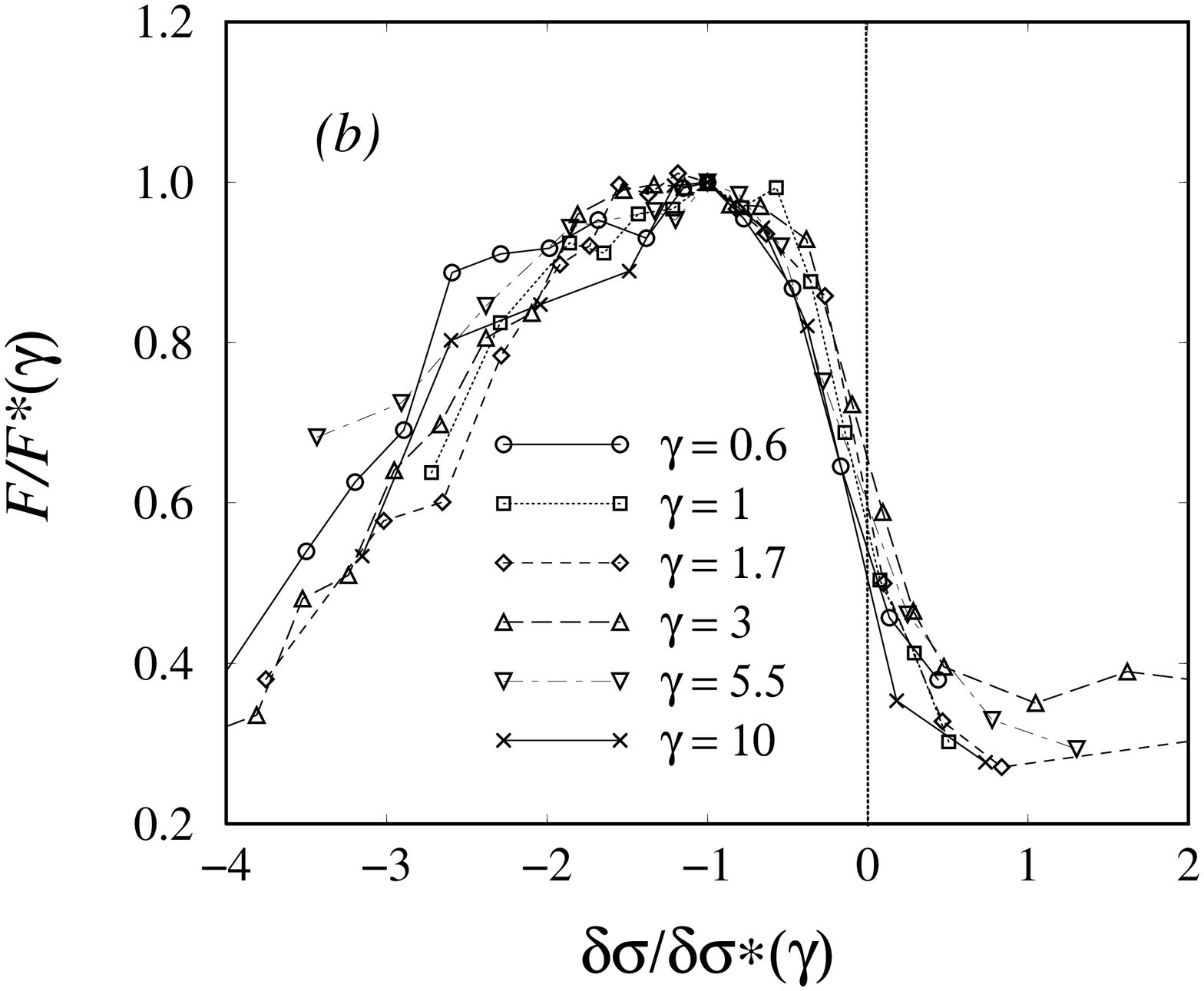, height=3.1in, angle=270}
\end{center}
\vspace{\baselineskip}
{\small Fig. 3: Fidelity of local alignments for piecewise correlated
sequences.
(a) The average fidelity $\ol \cF$ of the optimal alignment for various values
of the
scoring scoring parameters,
each averaged over an ensemble of $100$ --- $800$ sequences pairs.
The sequences are of length $N/2 = 10000$; they contain mutually
correlated subsequences of length $\hat N/2 = 2000$, which are related
by  Markovian evolution rules~\cite{Drasdo98} with parameters
$U = 0.3$ and $q = 0.25$.
(b)~The scaled curves $ \ol \cF / \cF^*(\gamma)$ as functions of
the scaled abscissa $x = \ds/|\ds^*(\gamma)|$ collapse to the single scaling
function $\varphi ( x^{3/2}  )$ in
accordance with Eqs.~(\ref{tsSsat}) and~(\ref{scalcor}).
}
\end{figure}

\begin{figure}
\begin{center}
\vspace*{-0.4cm}
\epsfig{file=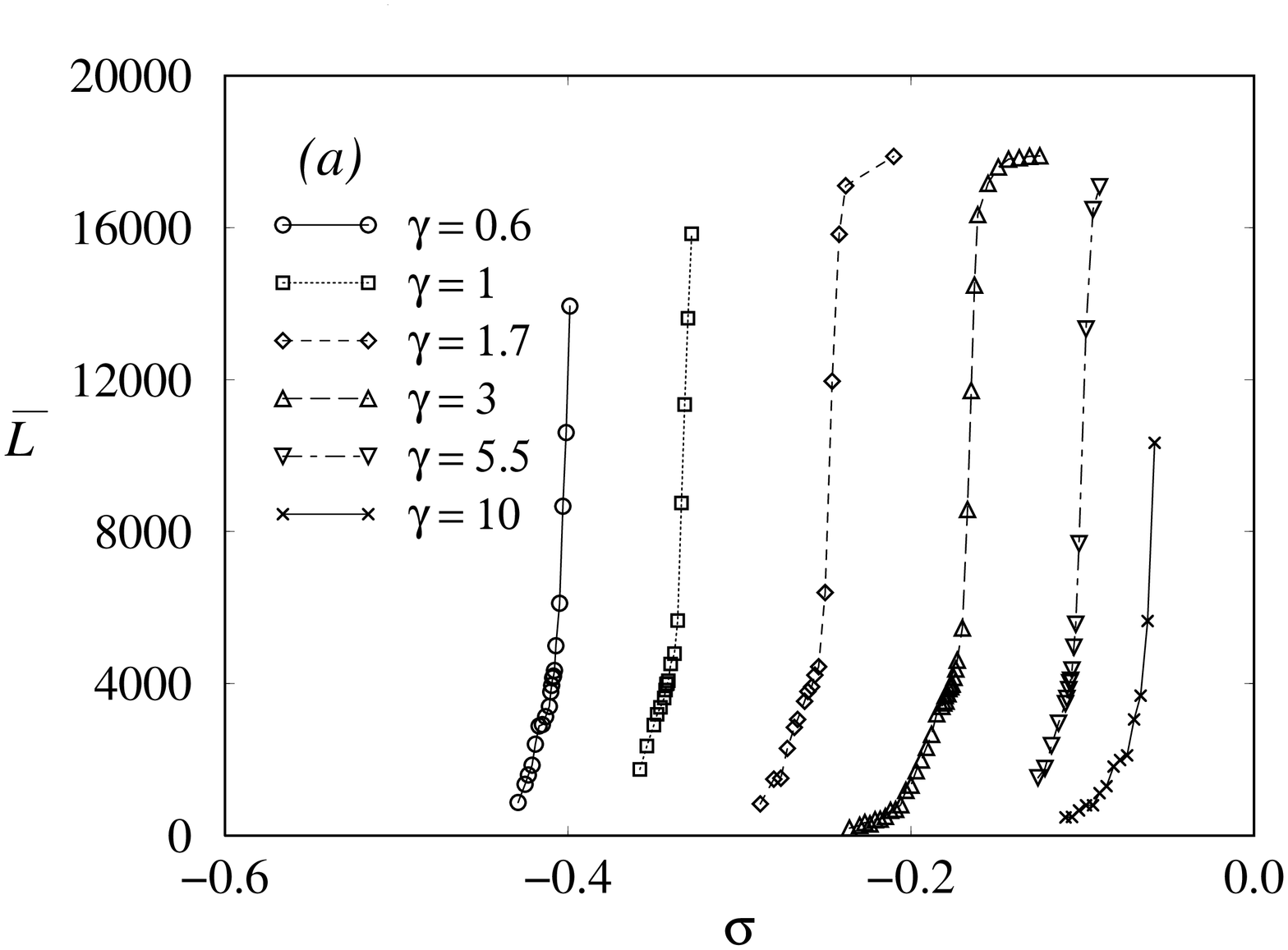, height=3.0in, angle=270}
\epsfig{file=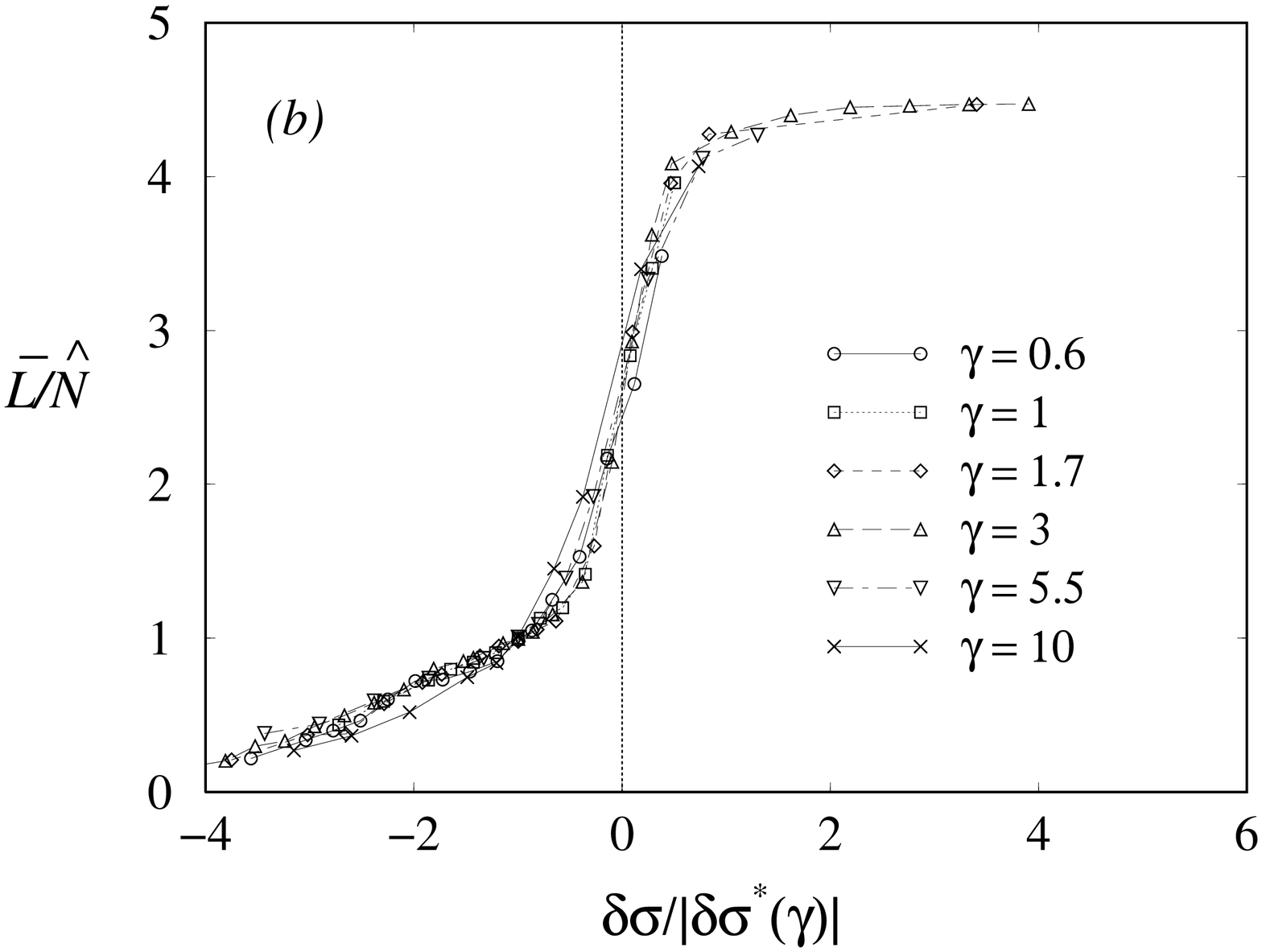, height=3.0in, angle=270}
\end{center}
\vspace{\baselineskip}
{\small Fig. 4: Length of local alignments for piecewise correlated sequences.
(a) The
average
length $\ol L$ of the optimal alignment for various
scoring parameters, obtained from the sequence pairs of Fig.~3.
(b)~The scaled curves $\ol  L / \hat N$ as functions of
the scaled abscissa $x = \ds/|\ds^*(\gamma)|$ collapse to the single scaling
function ${\cal L} ( x^{3/2}  )$ in
accordance with Eqs.~(\ref{tsSsat}) and~(\ref{scalcor}).
Note that at the point of maximal fidelity
($x = -1$), the alignment length equals the target length,
i.e., $ \ol L/\hat N \approx 1$. }
\end{figure}

\begin{figure}
\begin{center}
\epsfig{file=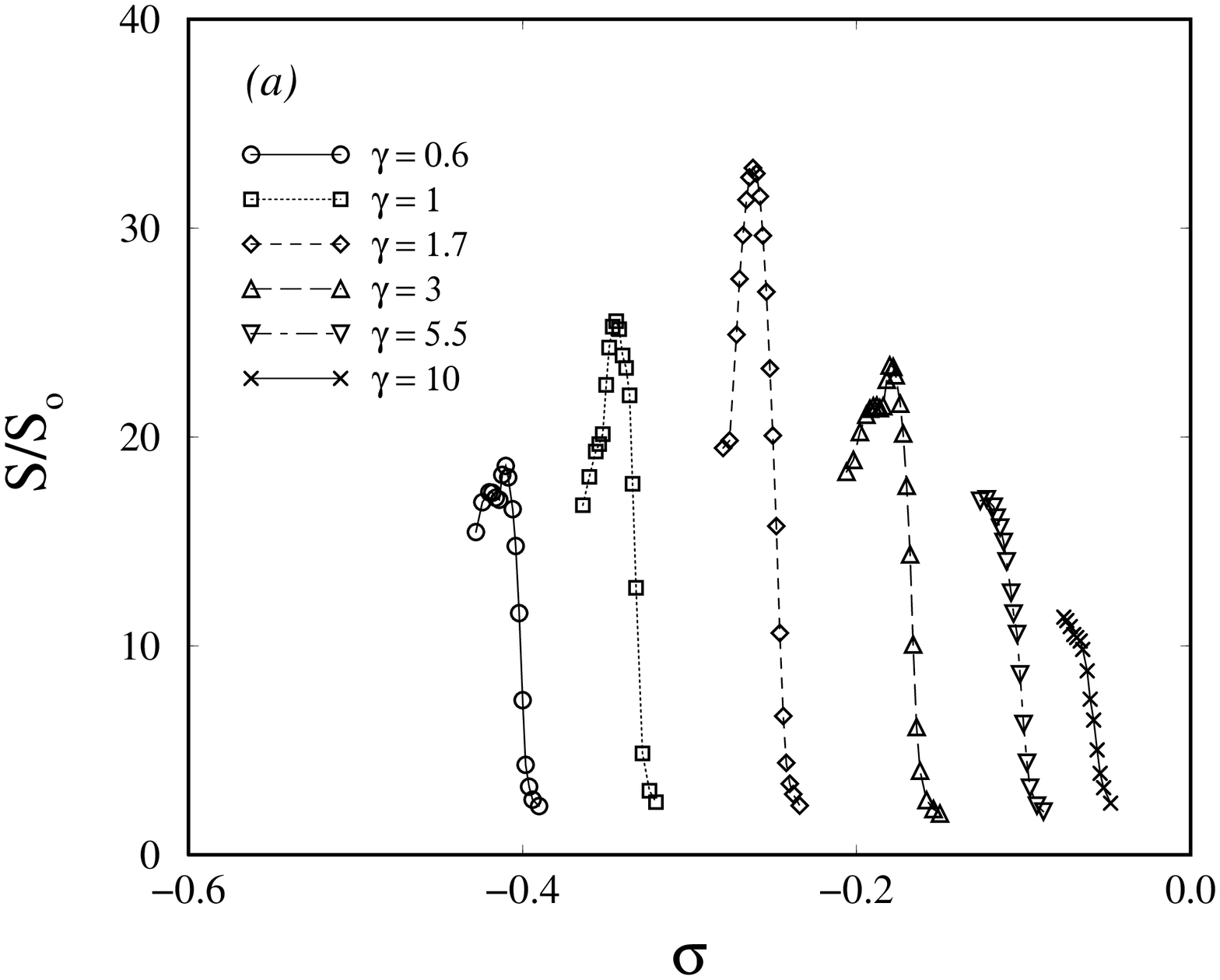, height=3.0in, angle=270}
\epsfig{file=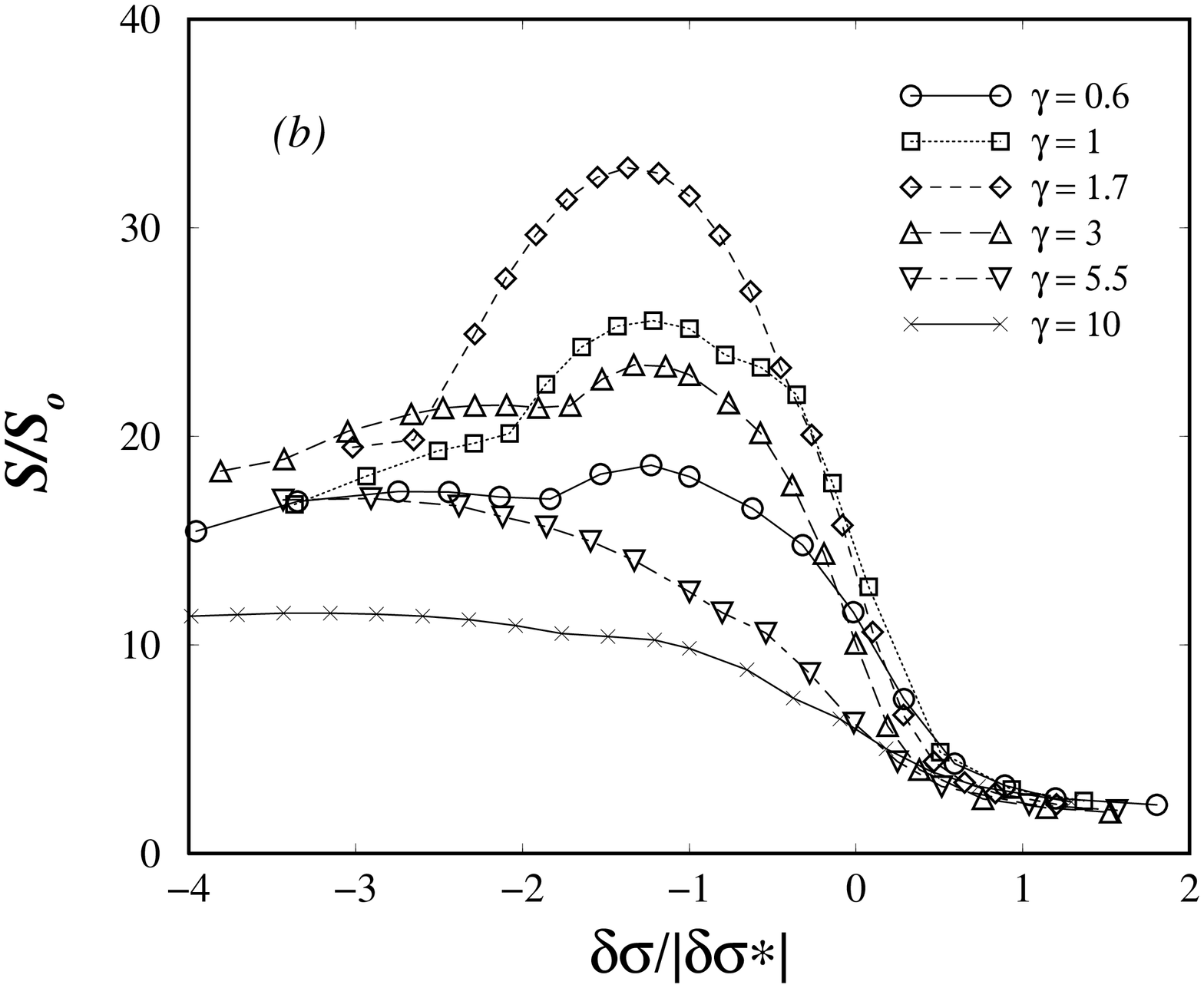, height=3.0in, angle=270}
\end{center}
\vspace{\baselineskip}
{\small Fig. 5: Score of local alignments for piecewise correlated sequences.
(a) The score ratio $s = S/S_0$ of the optimal alignment for various
scoring parameters, obtained from a {\em single} pair of the
 correlated sequences described in Fig.~3.
(b)~The scaled curves of $s$ as functions of
$x = \ds/|\ds^*(\gamma)|$ have maxima in the high-fidelity region
around the point $\ds / |\ds^*(\gamma)| = -1$; cf.~Fig.~3(b).  }
\end{figure}

\section{Parameter dependence and optimization}

As the fidelity curves of Fig.~3(a) show, the quality of an alignment
depends on the proper choice of both scoring parameters $\sigma$ and $\gamma$.
The strong
dependence of $\ol \cF$ on $\sigma$ can be understood by comparison
with Fig.~4. The relative fidelity maximum $\cF^*(\gamma)$ occurs at
a value $\ds^* (\gamma) < 0$ where the optimal alignment just covers the
target (i.e., $\ol L = \hat N$). For $\ds < \ds^*$, the optimal
alignment is too short. For $\ds > \ds^*$, the alignment ``overshoots''
the target, adding random matches to both sides and reducing its fidelity.
As $\ds \nearrow 0$, the length $\ol L$ increases continuously to values
of order $N$; that is, the optimal alignment becomes global.
Our result that $\overline{L} \approx \hat{N}$ when $\ds=\ds^*$
(Fig.~4) justifies the use of Eq.~(\ref{fidelity}) as a fidelity measure for 
local alignment.

For real alignment applications with unknown sequence correlations, 
the fidelity is of course not accessible
directly. What is readily accessible is the optimal score $S$ of an alignment. 
Below, we describe how the fidelity maximum can be {\em inferred} from
the score data. 
The key quantity to consider is the parameter dependence of  the {\em score ratio}
\begin{equation}
s(\sigma,\gamma) \equiv S / S_0 \;.
\label{s}
\end{equation}
As shown in Fig.~5 for alignment of a single pair of sequences, $s$ attains its 
relative maximum for fixed $\gamma$
at a value $\ds^{\rm max} (\gamma)$  close to $\ds^* (\gamma)$ and 
its absolute maximum at a value 
$\gamma^{\rm max} \approx \gamma^*$. 
More importantly, a comparison of Fig.~5(b) with Fig.~3(b) shows that 
{\em the fidelity 
$\cF(\sigma^{\rm max}, \gamma^{\rm max})$ evaluated at the maximum of $s$ 
is very close to the actual fidelity maximum $\cF^*$}.
 While the fidelity and score patterns fluctuate
for individual sequence pairs, this relationship between their maxima
turns out to be remarkably robust.
Our results therefore suggest that high-fidelity alignments can be obtained by
maximizing the score ratio $s$. As can be seen from Fig.~5, the
parameter dependence of $s(\sigma, \gamma)$ is given by a 
``mountain'' with a well-defined local maximum. The location of the maximum
$(\sigma^{\rm max},\gamma^{\rm max})$, and hence the location of the fidelity maximum, 
is accessible by a standard iterative procedure in a few steps.

This link between fidelity and score data is expected from
the scaling theory of alignment. For $\gamma \approx \gamma^*$,
the score ratio takes the scaling form
$\ol s = (\hat N / t_c) \, {\cal S} ( t_c / t_0  )$ similar to~(\ref{scalcor}).
The relative maxima $\cF^*(\gamma)$ and
$s^{\rm max} (\gamma) \equiv \max_\sigma \ol s(\sigma, \gamma)$ 
are determined by the maxima of the scaling functions $\varphi$ and $\cal S$,
respectively. These are functions of the same variable $\tau_c \equiv t_c /
t_0$; their maxima are found to occur at values $\tau_c^* \approx \tau_c^{\rm
max}$ both of order 1. The lines $\ds^*(\gamma)$ and $\ds^{\rm max} (\gamma)$
are then given by the equations $\tau_c (\sigma, \gamma) = \tau_c^*$ and
$\tau_c (\sigma, \gamma) = \tau_c^{\rm max}$, respectively.
The positions of the
absolute maxima turn out to be related in a similar way~\cite{Drasdo98}.
A more detailed discussion will be given elsewhere~\cite{Olsen98b}.

The optimization criterion can be reformulated in two ways:
\newline
(i) The relative maxima of the
score ratio define the function
$s^{\rm max} (\gamma) = \tau^{\rm max} N / t_c(\gamma)$. Hence, the global
maximum $s^{\rm max}$ is obtained by minimizing the threshold length $t_c$
while keeping $\tau \approx \tau^{\rm max}$, i.e., $t_0$ of order $t_c$.
\newline
(ii) The threshold length $t_c$ is related to another important
quantity, the score gain $\delta E$ over uncorrelated sequences
per aligned element in global alignments (see the detailed discussion
in Drasdo {\em et al.}~\cite{Drasdo98}).
We have $t_c \sim B^{3/2} (\gamma) (\delta E)^{-3/2}$.
By comparison with (\ref{tsSsat}), it follows that the above optimization
is equivalent to maximizing $\delta E$ while keeping $|\ds|$ of order
$\delta E$.

\section{Summary}

We have presented a conceptually simple and statistically well-founded
procedure to optimize Smith-Waterman alignments. For given scoring parameters
in the logarithmic phase, we compute (i) the score $S$ of the optimal
alignment and (ii) the average background score $S_0$ obtained by
randomizing the sequences. The scoring parameters are then improved
iteratively by maximizing the score ratio $S / S_0$.
We have shown that this procedure produces alignments of high fidelity on
test sequences.
Efficient algorithmic implementations and applications
to real biological sequences are currently being studied~\cite{Olsen98b}.

\section*{Acknowledgments}

The authors have benefited from discussions with S.F. Altschul, R. Bundschuh,
R. Durbin, M.A. Mu\~{n}oz, M. Vingron, and M.S. Waterman.
R.O. acknowledges a graduate fellowship from the La Jolla Interfaces in
Science Program supported by the Wellcome-Burroughs Foundation.
T.H. acknowledges  a
research fellowship by the A.P. Sloan Foundation, and an young investigator
award from the Arnold and Mabel Beckman Foundation.

\end{document}